# Fault tolerant channel-encrypting quantum dialogue against collective noise

Tian-Yu Ye*

College of Information & Electronic Engineering, Zhejiang Gongshang University, Hangzhou 310018, P.R.China

In this paper, two fault tolerant channel-encrypting quantum dialogue (QD) protocols against collective noise are presented. One is against collective-dephasing noise, while the other is against collective-rotation noise. The decoherent-free states, each of which is composed of two physical qubits, act as traveling states combating collective noise. Einstein-Podolsky-Rosen pairs, which play the role of private quantum key, are securely shared between two participants over a collective-noise channel in advance. Through encryption and decryption with private quantum key, the initial state of each traveling two-photon logical qubit is privately shared between two participants. Due to quantum encryption sharing of the initial state of each traveling logical qubit, the issue of information leakage is overcome. The private quantum key can be repeatedly used after rotation as long as the rotation angle is properly chosen, making quantum resource economized. As a result, their information-theoretical efficiency is nearly up to 66.7% . The proposed QD protocols only need single-photon measurements rather than two-photon joint measurements for quantum measurements. Security analysis shows that an eavesdropper cannot obtain anything useful about secret messages during the dialogue process without being discovered. Furthermore, the proposed QD protocols can be implemented with current techniques in experiment.
**Quantum dialogue(QD); quantum encryption; information leakage; decoherent-free(DF) state; collective-dephasing noise; collective-rotation noise**

## 1 Introduction

Quantum cryptography, whose security is based on the property of quantum mechanics rather than the computation difficulty of solving mathematical problems, can be regarded as the generalization of classical cryptography into the realm of quantum mechanics. It was first derived from the pioneering work of Bennett and Brassard [1] in 1984. At present, it has been expanded into many interesting branches, such as quantum key distribution (QKD) [1-8], quantum secret sharing (QSS) [9-13], quantum secure direct communication (QSDC) [14-26], and so on. On the aspect of QKD, many good works have been accomplished up to now. In 1991, Ekert [2] proposed a QKD protocol based on Bell's theorem with Einstein-Podolsky-Rosen (EPR) pairs. In 1992, Bennett et al. [3] simplified the security check process of Ekert's protocol. In 2000, Cabello [4] proposed a QKD protocol based on polarized photons with its efficiency equal to 100%. In 2003, Deng and Long [5] put forward a QKD protocol through the controlled order rearrangement encryption of the particles in photon pairs. In 2004, Deng and Long [6] suggested a bidirectional QKD protocol using practical faint laser pulses. In 2014, Su [7] proved that the Gaussian quantum discord state of optical field can be used to accomplish continuous variable quantum key distribution. In the same year, Zhang et al. [8] suggested a delayed error verification scheme for QKD using extra keys from privacy amplification with an arbitrarily small failure probability. On the aspect of QSS, Hillery et al. [9] proposed the first QSS protocol using a three-particle Greenberger-Horne-Zeilinger state in 1999. In the same year, Karlsson et al. [10] constructed a QSS protocol based on two-particle quantum entanglement. In 2004, Xiao et al. [11] put forward a common framework for multiparty QSS and provided two efficient ways for sharing a private key among three parties. In 2010, Hao et al. [12] designed a secure one-to-two-party QSS protocol with large capacity through Grover's algorithm in 16 randomly chosen quantum states. In 2011, Hao et al. [13] proposed a three-party QSS protocol through four-state Grover algorithm and implemented it with experiment successfully. It is well known that QSDC is able to transmit confidential messages directly between remote participants without establishing a sequence of random key first. In 2002, Long and Liu [14] proposed the first QSDC protocol, that is, the state-encoded two-step QSDC protocol. In this protocol, EPR pairs are used to represent secret messages and quantum data block transmission is first introduced to transmit quantum states from the sender to the receiver. In the same year, Bostrom and Felbinger [15] proposed the well-known ping-pong protocol with EPR pairs. In 2003, Deng et al. [16] suggested the operation-encoded two-step QSDC protocol. In this protocol, EPR pairs are encoded with unitary operations according to secret messages, and they are sent from the sender to the receiver in the manner of quantum data block transmission in two steps. In 2004, Deng and Long [17] proposed the quantum one-time pad (OTP) QSDC protocol. In 2005, Wang et al. [18] designed a QSDC protocol using high-dimension quantum superdense coding. In the same year, Wang et al. [19] suggested a multi-step QSDC protocol with multi-particle GHZ state. In 2008, Chen et al. [20] gave out a three-party controlled QSDC protocol with W state. In 2011, Gu et al. [21] proposed a two-step QSDC protocol with hyperentanglement. In 2012, Liu et al. [22] suggested a high-capacity QSDC protocol based on single photons in both the polarization and the spatial-mode degrees of freedom. In the same year, Sun et al. [23] constructed a QSDC protocol based on two-photon four-qubit cluster states. In 2013, Ren et al. [24] conducted photonic spatial Bell-state analysis for robust QSDC with quantum dot-cavity systems. In 2014, Zou and Qiu [25] designed a three-step semiquantum secure direct communication protocol with single photons in which the sender Alice is classical. In the same year, Chang et al. [26] constructed a controlled QSDC protocol with authentication through quantum OTP using five-particle cluster state.

Although QSDC has obtained considerable developments at present, most of QSDC protocols [14-26] just realize one-way communication, that is, they cannot accomplish the mutual exchange of secret bits from different participants. In 2004, Zhang et al. [27-28] and Nguyen [29] successfully made up for this drawback when they independently put forward the novel concept of quantum dialogue (QD). However, those earliest QD protocols [27-39] always run the risk of information leakage, which was first discovered by Gao et al. [40-41] in 2008, due to the existence of "classical correlation" [42]. It has been a common knowledge that the key to solve the issue of information leakage in QD lies in sharing the initial quantum state privately among participants. Several methods toward it have been suggested, such as the direct transmission of auxiliary quantum state [43-46], the correlation extractability of Bell states [47], the measurement correlation from entanglement swapping of two Bell states [48], quantum encryption sharing [49], and twice QSDC transmissions [50].

---

*Corresponding author:
E-mail：happyyty@aliyun.com(T.Y.Ye)

Accompanying the development of quantum cryptography, a special concept called channel-encrypting (or quantum-encrypting) quantum cryptography was derived. In a channel-encrypting quantum cryptography protocol [51-60], different participants usually share a sequence of quantum states as their reusable private quantum key first, and then use it to encrypt and decrypt the traveling states carrying classical secret bits successively with some operations. For example, in the channel-encrypting quantum cryptography protocols of Refs. [51,59], EPR pairs are used as reusable private quantum key for encrypting and decrypting traveling particles with controlled-NOT (CNOT) operation. In the multiparty channel-encrypting QSDC protocols of Refs. [55-56], multi-particle GHZ states act as reusable private quantum key for encrypting and decrypting traveling particles with CNOT operation. In the channel-encrypting QSDC protocol of Ref. [57], two-photon pure entangled states are used as reusable private quantum key for encrypting and decrypting traveling particles with CNOT operation. Apparently, the initial traveling states can be easily shared between two participants through quantum encryption. In Ref. [49], the idea of quantum encryption was introduced into QD by the author to share the initial quantum state privately between two participants.

In a practical transmission, the fluctuation of the birefringence of optical fiber alters the polarization of photons, resulting in channel noise. At present, lots of quantum cryptography protocols are merely workable under the assumption of an ideal channel, such as those QD protocols in Refs. [27-39,43-50]. Apparently, how to make quantum cryptography protocols work well over a noisy channel is of great significance. There has emerged several good methods easing the effect of noise, such as entanglement purification [61-67], quantum error correct code [68], single-photon error rejection [69], and decoherence-free(DF) states [70-80]. The drawback of the former three strategies is that they only take effect under the following assumptions: the interaction between photons and the environment is weak enough and photons are disturbed with a low probability. Fortunately, channel noise can be modeled as collective noise, which means that if several photons transmit over a noisy channel simultaneously or they are spatially close, the transformation of noise on each photon is identical. [73-74] Because the DF states are invariant toward collective noise, they are frequently used to defeat this kind of noise.

In this paper, two fault tolerant channel-encrypting QD protocols against collective noise are presented. One is against collective-dephasing noise, while the other is against collective-rotation noise. The DF states, each of which is composed of two physical qubits, act as traveling states combating collective noise. EPR pairs, which play the role of private quantum key, are securely shared between two participants over a collective-noise channel in advance. Through encryption and decryption with private quantum key, the initial state of each traveling two-photon logical qubit is privately shared between two participants. Due to quantum encryption sharing of the initial state of each traveling logical qubit, the issue of information leakage is overcome. The private quantum key can be repeatedly used after rotation as long as the rotation angle is properly chosen, making quantum resource economized. As a result, their information-theoretical efficiency is nearly up to 66.7%. The proposed QD protocols only need single-photon measurements rather than two-photon joint measurements for quantum measurements. Security analysis shows that an eavesdropper cannot obtain anything useful about secret messages during the dialogue process without being discovered. Furthermore, the proposed QD protocols can be implemented with current techniques in experiment.

## 2 Fault tolerant channel-encrypting QD protocols against collective noise
### 2.1 Fault tolerant channel-encrypting QD protocol against collective-dephasing noise

The collective-dephasing noise over a quantum channel can be depicted as [70]

$$U_{dp}|0\rangle = |0\rangle, U_{dp}|1\rangle = e^{i\varphi}|1\rangle, \tag{1}$$

where $\varphi$ is the noise parameter changing along time, and $|0\rangle$ and $|1\rangle$ are the horizontal and the vertical polarizations of photons, respectively. The two logical qubits [70],

$$|0\rangle_{dp} \equiv |01\rangle, |1\rangle_{dp} \equiv |10\rangle, \tag{2}$$

each of which is composed of two physical qubits with an antiparallel parity, are invariant toward this kind of noise. The superpositions of these two logical qubits, that is, [72]

$$|\pm\rangle_{dp} = \frac{1}{\sqrt{2}}\left(|0\rangle_{dp} \pm |1\rangle_{dp}\right) \equiv \frac{1}{\sqrt{2}}\left(|01\rangle \pm |10\rangle\right) = |\psi^{\pm}\rangle, \tag{3}$$

are also invariant toward this kind of noise. We define two logical unitary operations as [74]

$$U_0^{dp} = I_1 \otimes I_2, U_1^{dp} = \left(-i\sigma_y\right)_1 \otimes \left(\sigma_x\right)_2, \tag{4}$$

where $I = |0\rangle\langle 0| + |1\rangle\langle 1|$, $-i\sigma_y = |1\rangle\langle 0| - |0\rangle\langle 1|$, and $\sigma_x = |1\rangle\langle 0| + |0\rangle\langle 1|$ are three general unitary operations, and the subscripts 1 and

2 in three unitary operations denote the first and the second physical qubits in each logical qubit, respectively. After simple deduction, it has [77]

$$U_1^{dp}|0\rangle_{dp}=|1\rangle_{dp}, U_1^{dp}|1\rangle_{dp}=-|0\rangle_{dp}, U_1^{dp}|+\rangle_{dp}=-|-\rangle_{dp}, U_1^{dp}|-\rangle_{dp}=|+\rangle_{dp}. \quad (5)$$

We assume that Alice has $N$ bits secret messages consisting of $\{j_1, j_2, \cdots, j_N\}$, and Bob has $N$ bits consisting of $\{k_1, k_2, \cdots, k_N\}$, where $j_i, k_i \in \{0,1\}$ $(i=1,2,\cdots,N)$. The fault tolerant channel-encrypting QD protocol against collective-dephasing noise is made up of the following steps.

**Step 1:** Alice and Bob share $N$ EPR pairs $|\phi^+\rangle_{AB} = \frac{1}{\sqrt{2}}(|0\rangle_A|0\rangle_B + |1\rangle_A|1\rangle_B)$ as their private quantum key over a collective-dephasing noise channel by using the following method, which is derived from Ref. [74].

①Alice prepares $N+\delta_1$ entangled states

$$|\Theta_{dp}^+\rangle_{AC} = \frac{1}{\sqrt{2}}(|0\rangle|0\rangle_{dp} + |1\rangle|1\rangle_{dp})_{AC} = \frac{1}{\sqrt{2}}(|0\rangle_A|01\rangle_{C_1C_2} + |1\rangle_A|10\rangle_{C_1C_2})$$
$$= \frac{1}{2}\left[|+\rangle_A(|++\rangle-|--\rangle)_{C_1C_2} + |-\rangle_A(|-+\rangle-|+-\rangle)_{C_1C_2}\right]. \quad (6)$$

She divides these entangled states into two photon sequences, $S_A$ and $S_C$, where $S_A$ is composed of all the photons $A$ and $S_C$ is composed of all the logical qubits $C$. Afterward, she keeps $S_A$ by herself and sends $S_C$ to Bob in the manner of quantum data block transmission [14].

②After Bob announces the receipt of $S_C$, they implement the security check procedure together using the sampling check method similar to the one in Ref. [16]. Bob randomly picks out $\delta_1$ logical qubits from $S_C$ and measures each sample logical qubit randomly with one of the two bases $\sigma_z \otimes \sigma_z$ and $\sigma_x \otimes \sigma_x$, where $\sigma_z = \{|0\rangle, |1\rangle\}$ and $\sigma_x = \{|+\rangle, |-\rangle\}$. Then, he tells Alice the positions and the measurement bases of these sample logical qubits. Alice measures the corresponding sample photons $A$ in $S_A$ with the proper measurement bases. That is, if Bob uses the base $\sigma_z \otimes \sigma_z$ ($\sigma_x \otimes \sigma_x$) to measure a sample logical qubit $C$ in $S_C$, Alice will choose the base $\sigma_z$ ($\sigma_x$) to measure the corresponding sample photon $A$ in $S_A$. After Alice publishes her measurement outcomes, the existence of an eavesdropper can be judged out by Bob through the deterministic entanglement correlations shown in Eq.(6). As long as the transmission security is guaranteed, they can successfully share the remaining $N$ entangled states $|\Theta_{dp}^+\rangle_{AC}$.

③For each of the remaining $N$ entangled states $|\Theta_{dp}^+\rangle_{AC}$, Bob performs a CNOT operation on photons $C_1$ and $C_2$ ($C_1$ is the control qubit and $C_2$ is the target qubit) [60]. As a result, $|\Theta_{dp}^+\rangle_{AC}$ is changed into

$$|\Xi_{dp}\rangle_{AC} = \frac{1}{\sqrt{2}}(|0\rangle_A|01\rangle_{C_1C_2} + |1\rangle_A|11\rangle_{C_1C_2}) = \frac{1}{\sqrt{2}}(|0\rangle_A|0\rangle_{C_1} + |1\rangle_A|1\rangle_{C_1})|1\rangle_{C_2}. \quad (7)$$

Now Alice and Bob successfully share $N$ EPR pairs $|\phi^+\rangle_{AC_1} = \frac{1}{\sqrt{2}}(|0\rangle_A|0\rangle_{C_1} + |1\rangle_A|1\rangle_{C_1})$. Without loss of generality, the subscript $C_1$ in $|\phi^+\rangle_{AC_1}$ can be replaced by $B$.

**Step 2:** Alice prepares a sequence of $N$ traveling states $\{|m_1\rangle_{dp}, |m_2\rangle_{dp}, \cdots, |m_N\rangle_{dp}\}$, where $m_i = 0$ or $1$ $(i=1,2,\cdots,N)$. For convenience, this sequence is called as $S_M$. To guarantee the transmission security, the usage of decoy photons [81-82] is adopted by Alice. That is, she prepares some decoy photons randomly in one of the four states $\{|0\rangle_{dp}, |1\rangle_{dp}, |+\rangle_{dp}, |-\rangle_{dp}\}$ and inserts them randomly into $S_M$. As a result, $S_M$ is turned into $S_M'$. Alice uses private quantum key $|\phi^+\rangle_{AB}$ to encrypt the traveling states in $S_M'$ except for the decoy photons. That is, Alice performs a controlled-$U_1^{dp}$ ($CU_1^{dp}$) operation [60] on the photons $A_i$ and

$\left|m_i\right\rangle_{dp}$ ($A_i$ is the controller and $\left|m_i\right\rangle_{dp}$ is the target), where

$$CU_1^{dp} = |0\rangle\langle 0| \otimes U_0^{dp} + |1\rangle\langle 1| \otimes U_1^{dp}. \tag{8}$$

Afterward, Alice sends $S_M^{'}$ to Bob through block transmission [14] and tells Bob the positions and the preparation bases of decoy photons in $S_M^{'}$ when Bob announces its receipt. Then, Bob tells Alice his measurement outcomes after measuring the decoy photons with the bases Alice told. The existence of an eavesdropper can be judged out by Alice through the consistency between the initial states of decoy photons and Bob's measurement outcomes on them. If there is no eavesdropper, they go on the next step; otherwise, they restart from the beginning.

**Step 3:** Bob drops out the decoy photons in $S_M^{'}$ to get $S_M$. Then, Bob decrypts out the traveling states in $S_M$. That is, Bob performs a $CU_1^{dp}$ operation on the photons $B_i$ and $\left|m_i\right\rangle_{dp}$ ($B_i$ is the controller and $\left|m_i\right\rangle_{dp}$ is the target). Bob measures $\left|m_i\right\rangle_{dp}$ with the base $\sigma_z \otimes \sigma_z$ to know its initial state. According to his measurement outcome, Bob reproduces a new traveling state $\left|m_i\right\rangle_{dp}$ with no measurement performed. For encoding his one-bit secret message $k_i$, Bob performs the logical unitary operation $U_{k_i}^{dp}$ on the new $\left|m_i\right\rangle_{dp}$, resulting in $U_{k_i}^{dp}\left|m_i\right\rangle_{dp}$. To guarantee the transmission security, Bob also prepares some decoy photons randomly in one of the four states $\left\{\left|0\right\rangle_{dp}, \left|1\right\rangle_{dp}, \left|+\right\rangle_{dp}, \left|-\right\rangle_{dp}\right\}$ and inserts them randomly into $S_M$. As a result, $S_M$ is turned into $S_M^{''}$. Then, Bob sends $S_M^{''}$ to Alice through block transmission [14]. After Alice announces the receipt of $S_M^{''}$, they implement the security check procedure similar to that of Step 2. If the transmission security of $S_M^{''}$ is guaranteed, they go on the next step.

**Step 4:** Alice drops out the decoy photons in $S_M^{''}$ to get $S_M$. For encoding her one-bit secret message $j_i$, Alice performs the logical unitary operation $U_{j_i}^{dp}$ on $U_{k_i}^{dp}\left|m_i\right\rangle_{dp}$ in $S_M$. Accordingly, $U_{k_i}^{dp}\left|m_i\right\rangle_{dp}$ is turned into $U_{j_i}^{dp}U_{k_i}^{dp}\left|m_i\right\rangle_{dp}$. Afterward, Alice measures $U_{j_i}^{dp}U_{k_i}^{dp}\left|m_i\right\rangle_{dp}$ with the base $\sigma_z \otimes \sigma_z$. For bidirectional communication, Alice announces the measurement outcome of $U_{j_i}^{dp}U_{k_i}^{dp}\left|m_i\right\rangle_{dp}$ publicly. According to $U_{j_i}^{dp}$ and the measurement outcome of $U_{j_i}^{dp}U_{k_i}^{dp}\left|m_i\right\rangle_{dp}$, Alice can read out $k_i$, since she prepares $\left|m_i\right\rangle_{dp}$ by herself. Likewise, Bob can also read out $j_i$, according to $U_{k_i}^{dp}$ and the measurement outcome of $U_{j_i}^{dp}U_{k_i}^{dp}\left|m_i\right\rangle_{dp}$. For example, if $\left|m_1\right\rangle_{dp} = \left|1\right\rangle_{dp}$, $j_1 = 1$, and $k_1 = 0$, after both of Alice and Bob's encoding operations, the final encoded state of $\left|m_1\right\rangle_{dp}$ (i.e., $U_{j_1}^{dp}U_{k_1}^{dp}\left|m_1\right\rangle_{dp}$) will be $\left|0\right\rangle_{dp}$. Then, from $U_{j_1}^{dp}$ and her own measurement outcome of $U_{j_1}^{dp}U_{k_1}^{dp}\left|m_1\right\rangle_{dp}$, Alice can easily decode out that $k_1 = 0$. Similar thing also happens to Bob.

**Step 5:** For each of their private quantum key $\left|\phi^+\right\rangle_{AB}$, Alice and Bob perform the rotation operation depicted as Eq.(9) on their respective photon by choosing a proper angle $\theta$, respectively [51]:

$$R(\theta) = \begin{pmatrix} \cos\theta & \sin\theta \\ -\sin\theta & \cos\theta \end{pmatrix}. \tag{9}$$

To prevent an eavesdropping behavior from an outside eavesdropper effectively, $\theta$ should satisfy that $\theta \neq k\pi \pm \pi/4$ ($k = 0, \pm 1, \pm 2, \cdots$). [54] After the rotation, Alice and Bob can reuse EPR pairs $\left|\phi^+\right\rangle_{AB}$ as their private quantum key when restarting their next round communication from Step 2.

In fact, only single-photon measurements rather than two-photon joint measurements are needed for quantum measurements in the proposed QD protocol. The reason is that, after a Hadamard operation performed on each physical qubit, $\left|+\right\rangle_{dp}$ is turned into $\left|\phi^-\right\rangle = \frac{1}{\sqrt{2}}(|0\rangle|0\rangle - |1\rangle|1\rangle)$, while $\left|-\right\rangle_{dp} = \left|\psi^-\right\rangle = \frac{1}{\sqrt{2}}(|0\rangle|1\rangle - |1\rangle|0\rangle)$ is kept unchanged. [77] These two Bell states can be easily discriminated through two single-photon measurements, as the parity of two photons is parallel for $\left|\phi^-\right\rangle$ and antiparallel for $\left|\psi^-\right\rangle$. Therefore, when Alice or Bob needs to discriminate between $\left|+\right\rangle_{dp}$ and $\left|-\right\rangle_{dp}$, she or he only needs two single-photon measurements after performing a Hadamard operation on each physical qubit.

## 2.2 Fault tolerant channel-encrypting QD protocol against collective-rotation noise

The collective-rotation noise over a quantum channel can be depicted as [72]

$$U_r|0\rangle = \cos\theta|0\rangle + \sin\theta|1\rangle,$$
$$U_r|1\rangle = -\sin\theta|0\rangle + \cos\theta|1\rangle, \tag{10}$$

where $\theta$ is the noise parameter changing along time. Apparently, the following two logical qubits are invariant toward this kind of noise [72]:

$$|0\rangle_r = |\phi^+\rangle, \ |1\rangle_r = |\psi^-\rangle, \tag{11}$$

where $|\phi^+\rangle = \frac{1}{\sqrt{2}}(|00\rangle + |11\rangle)$. The superpositions of these two logical qubits, that is, [76-77]

$$|+\rangle_r = \frac{1}{\sqrt{2}}(|0\rangle_r + |1\rangle_r), |-\rangle_r = \frac{1}{\sqrt{2}}(|0\rangle_r - |1\rangle_r), \tag{12}$$

are also invariant toward this kind of noise. After simple deduction, it has [77]

$$U_1^r|0\rangle_r = |1\rangle_r, \ U_1^r|1\rangle_r = -|0\rangle_r, \ U_1^r|+\rangle_r = -|-\rangle_r, \ U_1^r|-\rangle_r = |+\rangle_r, \tag{13}$$

where two logical unitary operations are [74]

$$U_0^r = I_1 \otimes I_2, \ U_1^r = I_1 \otimes (-i\sigma_y)_2. \tag{14}$$

We assume that Alice has $N$ bits secret messages consisting of $\{j_1, j_2, \cdots, j_N\}$, and Bob has $N$ bits consisting of $\{k_1, k_2, \cdots, k_N\}$, where $j_i, k_i \in \{0,1\}$ $(i = 1, 2, \cdots, N)$. The fault tolerant channel-encrypting QD protocol against collective-rotation noise is made up of the following steps.

**Step 1:** Alice and Bob share $N$ EPR pairs $|\phi^+\rangle_{AB} = \frac{1}{\sqrt{2}}(|0\rangle_A|0\rangle_B + |1\rangle_A|1\rangle_B)$ as their private quantum key over a collective-rotation noise channel by using the following method, which is derived from Ref. [74].

①Alice prepares $N + \delta_1$ entangled states

$$|\Theta_r^+\rangle_{AC} = \frac{1}{\sqrt{2}}(|0\rangle|0\rangle_r + |1\rangle|1\rangle_r)_{AC} = \frac{1}{2}\left[|0\rangle_A(|00\rangle + |11\rangle)_{C_1C_2} + |1\rangle_A(|01\rangle - |10\rangle)_{C_1C_2}\right]$$
$$= \frac{1}{\sqrt{2}}(|+y\rangle_A|+y\rangle_{C_1}|-y\rangle_{C_2} + |-y\rangle_A|-y\rangle_{C_1}|+y\rangle_{C_2}), \tag{15}$$

where

$$|+y\rangle = \frac{1}{\sqrt{2}}(|0\rangle + i|1\rangle), |-y\rangle = \frac{1}{\sqrt{2}}(|0\rangle - i|1\rangle). \tag{16}$$

She divides these entangled states into two photon sequences, $S_A$ and $S_C$, where $S_A$ is composed of all the photons $A$ and $S_C$ is composed of all the logical qubits $C$. Afterward, she keeps $S_A$ by herself and sends $S_C$ to Bob through block transmission [14].

②After Bob announces the receipt of $S_C$, they implement the security check procedure together using the sampling check method similar to the one in Ref. [16]. Bob randomly picks out $\delta_1$ logical qubits from $S_C$ and measures each sample logical qubit randomly with one of the two bases $\sigma_z \otimes \sigma_z$ and $\sigma_y \otimes \sigma_y$, where $\sigma_y = \{|+y\rangle, |-y\rangle\}$. He tells Alice the positions and the measurement bases of these sample logical qubits. Then, Alice measures the corresponding sample photons $A$ in $S_A$ with the proper measurement bases. That is, if Bob uses the base $\sigma_z \otimes \sigma_z$ ($\sigma_y \otimes \sigma_y$) to measure a sample logical qubit $C$ in $S_C$, Alice will choose the base $\sigma_z$ ($\sigma_y$) to measure the corresponding sample photon $A$ in $S_A$. After Alice publishes her measurement outcomes, the existence of an eavesdropper can be judged out by Bob through the deterministic entanglement correlations shown in Eq.(15). As long as the transmission security is guaranteed, they can successfully share the remaining $N$ entangled states $|\Theta_r^+\rangle_{AC}$.

③ Alice and Bob perform operations $S \otimes S \otimes S$ and $H \otimes H \otimes H$ on each of the remaining $N$ entangled states $|\Theta_r^+\rangle_{AC}$ together [60], where $S$ and $H$ are the phase gate and the Hadamard gate, respectively, that is,

$$S = \begin{pmatrix} 1 & 0 \\ 0 & i \end{pmatrix}, H = \frac{1}{\sqrt{2}} \begin{pmatrix} 1 & 1 \\ 1 & -1 \end{pmatrix}. \qquad (17)$$

As a result, $|\Theta_r^+\rangle_{AC}$ is changed into

$$|\Lambda_r\rangle_{AC} = \frac{1}{\sqrt{2}} \left( |1\rangle_A |1\rangle_{C_1} |0\rangle_{C_2} + |0\rangle_A |0\rangle_{C_1} |1\rangle_{C_2} \right). \qquad (18)$$

④ For each of $N$ entangled states $|\Lambda_r\rangle_{AC}$, Bob performs a CNOT operation on photons $C_1$ and $C_2$ ($C_1$ is the control qubit and $C_2$ is the target qubit) [60]. As a result, $|\Lambda_r\rangle_{AC}$ is changed into

$$|\Xi_r\rangle_{AC} = \frac{1}{\sqrt{2}} \left( |1\rangle_A |1\rangle_{C_1} |1\rangle_{C_2} + |0\rangle_A |0\rangle_{C_1} |1\rangle_{C_2} \right) = \frac{1}{\sqrt{2}} \left( |1\rangle_A |1\rangle_{C_1} + |0\rangle_A |0\rangle_{C_1} \right) |1\rangle_{C_2}. \qquad (19)$$

Now Alice and Bob successfully share $N$ EPR pairs $|\phi^+\rangle_{AC_1} = \frac{1}{\sqrt{2}} \left( |0\rangle_A |0\rangle_{C_1} + |1\rangle_A |1\rangle_{C_1} \right)$. Without loss of generality, the subscript $C_1$ in $|\phi^+\rangle_{AC_1}$ can be replaced by $B$.

**Step 2:** Alice prepares a sequence of $N$ traveling states $\{|m_1\rangle_r, |m_2\rangle_r, \cdots, |m_N\rangle_r\}$, where $m_i = 0$ or $1$ ($i = 1, 2, \cdots, N$). For convenience, this sequence is called as $S_M$. To guarantee the transmission security, the usage of decoy photons [81-82] is adopted by Alice. That is, Alice prepares some decoy photons randomly in one of the four states $\{|0\rangle_r, |1\rangle_r, |+\rangle_r, |-\rangle_r\}$ and inserts them randomly into $S_M$. As a result, $S_M$ is turned into $S_M'$. Alice uses private quantum key $|\phi^+\rangle_{AB}$ to encrypt the traveling states in $S_M'$ except for the decoy photons. That is, Alice performs a controlled-$U_1^r$ ($CU_1^r$) operation [60] on the photons $A_i$ and $|m_i\rangle_r$ ($A_i$ is the controller and $|m_i\rangle_r$ is the target), where

$$CU_1^r = |0\rangle\langle 0| \otimes U_0^r + |1\rangle\langle 1| \otimes U_1^r. \qquad (20)$$

Afterward, Alice sends $S_M'$ to Bob through block transmission [14]. After Bob announces the receipt of $S_M'$, they implement the security check procedure similar to that of Step 2 in Section 2.1.

**Step 3:** Bob drops out the decoy photons in $S_M'$ to get $S_M$. Then, Bob decrypts out the traveling states in $S_M$. That is, Bob performs a $CU_1^r$ operation on the photons $B_i$ and $|m_i\rangle_r$ ($B_i$ is the controller and $|m_i\rangle_r$ is the target). Then, Bob measures $|m_i\rangle_r$ with Bell basis to know its initial state. According to his measurement outcome, Bob reproduces a new traveling state $|m_i\rangle_r$ with no measurement performed. Then, Bob performs the logical unitary operation $U_{k_i}^r$ on the new $|m_i\rangle_r$ to encode his one-bit secret message $k_i$, resulting in $U_{k_i}^r |m_i\rangle_r$. To guarantee the transmission security, Bob also prepares some decoy photons randomly in one of the four states $\{|0\rangle_r, |1\rangle_r, |+\rangle_r, |-\rangle_r\}$ and inserts them randomly into $S_M$. As a result, $S_M$ is turned into $S_M''$. Then, Bob sends $S_M''$ to Alice through block transmission [14]. After Alice announces the receipt of $S_M''$, they implement the security check procedure similar to that of Step 2.

**Step 4:** Alice drops out the decoy photons in $S_M''$ to get $S_M$. For encoding her one-bit secret message $j_i$, Alice performs the logical unitary operation $U_{j_i}^r$ on $U_{k_i}^r |m_i\rangle_r$ in $S_M$. Accordingly, $U_{k_i}^r |m_i\rangle_r$ is turned into $U_{j_i}^r U_{k_i}^r |m_i\rangle_r$. Afterward, Alice measures $U_{j_i}^r U_{k_i}^r |m_i\rangle_r$ with Bell basis. For bidirectional communication, Alice announces the measurement outcome of $U_{j_i}^r U_{k_i}^r |m_i\rangle_r$ publicly. According to $U_{j_i}^r$ and the measurement outcome of $U_{j_i}^r U_{k_i}^r |m_i\rangle_r$, Alice can read out $k_i$, since she prepares $|m_i\rangle_r$ by herself. Likewise, Bob can also read out $j_i$, according to $U_{k_i}^r$ and the measurement outcome of $U_{j_i}^r U_{k_i}^r |m_i\rangle_r$.

**Step 5:** This step is the same as Step 5 in Section 2.1.

In fact, only single-photon measurements rather than two-photon joint measurements are needed for quantum measurements in the proposed QD protocol, similar to the case of collective-dephasing noise. A Hadamard operation performed on the second physical qubit can change $|+\rangle_r$ and $|-\rangle_r$ into $|\phi^-\rangle = \frac{1}{\sqrt{2}}(|0\rangle|0\rangle - |1\rangle|1\rangle)$ and $|\psi^+\rangle = \frac{1}{\sqrt{2}}(|0\rangle|1\rangle + |1\rangle|0\rangle)$, respectively [77]. Then, $|\phi^-\rangle$ and $|\psi^+\rangle$ can be easily discriminated through two single-photon measurements as they have opposite parities for their own two photons. Therefore, when Alice or Bob needs to discriminate between $|+\rangle_r$ and $|-\rangle_r$, she or he only needs two single-photon measurements after performing a Hadamard operation on the second physical qubit. On the other hand, $|\phi^+\rangle$ and $|\psi^-\rangle$ can also be easily discriminated by Alice or Bob through two single-photon measurements due to the opposite parities for their own two photons.

So far, the demonstration of the two fault tolerant channel-encrypting QD protocols against collective noise has been finished. The two proposed QD protocols work well over a collective-noise channel by exploiting the logical qubits to defeat collective noise, while the QD protocol in Ref. [49] only takes effect under an ideal channel. However, their basic principles for dialogue process are similar. Therefore, the two proposed QD protocols can be regarded as the generalization of the QD protocol in Ref. [49] under the case of collective noise.

## 3 Security analysis

As the bidirectional communication principles for the two proposed QD protocols are similar, without loss of generality, the author merely conducts security analysis on the one against collective-dephasing noise here.

(1) The information leakage problem

As for the example in Section 2.1, Bob is able to know the initial state of $|m_1\rangle_{dp}$ through private quantum key $|\phi^+\rangle_{A_1 B_1}$. Consequently, it is unnecessary for Alice to publish the initial state of $|m_1\rangle_{dp}$, making Eve unaware of it. Therefore, Alice's announcement on the measurement outcome of $U_{j_1}^{dp} U_{k_1}^{dp} |m_1\rangle_{dp}$ (i.e., $|0\rangle_{dp}$) means to Eve four possibilities of secret messages. That is, $\{j_1 = 0, k_1 = 0\}$, $\{j_1 = 1, k_1 = 1\}$, $\{j_1 = 1, k_1 = 0\}$, and $\{j_1 = 0, k_1 = 1\}$. According to Shannon's information theory [83], these involve $-\sum_{i=1}^{4} p_i \log_2 p_i = -4 \times \frac{1}{4} \log_2 \frac{1}{4} = 2$ bit information. Therefore, nothing about $\{j_1, k_1\}$ is leaked out to Eve.

(2) Eve's active attacks

The security analysis on Eve's active attacks is demonstrated according to the implementation steps.

**Step 1:** Alice prepares $N + \delta_1$ $|\Theta_{dp}^+\rangle_{AC}$ and sends $S_C$ to Bob. During the transmission, Eve may try to obtain $S_C$. If Eve is smart enough to perform the CNOT operation on photons $C_1$ and $C_2$, she can use $S_C$ to decrypt the ciphertext Alice sends to Bob later. As a result, Eve can know the initial state of $|m_i\rangle_{dp}$ $(i = 1, 2, \cdots, N)$. In this way, partial information of Alice and Bob's secret messages will be leaked out to Eve after Alice announces the measurement outcome of $U_{j_i}^{dp} U_{k_i}^{dp} |m_i\rangle_{dp}$ publicly. Fortunately, the following reasons make this thing impossible: (1) the quantum no-cloning theorem ensures that Eve is unable to replicate $S_C$ [59] and (2) any attack from Eve will be detected by Alice and Bob because the deterministic entanglement correlations between the sample logical qubits $C$ in $S_C$ and the corresponding sample photons $A$ in $S_A$ are destroyed. [16] In all, the private quantum key sharing process is secure in principle.

**Step 2:** Alice encrypts the traveling states and sends the ciphertext to Bob. The encryption on $|m_i\rangle_{dp}$ with a $CU_1^{dp}$ operation from Alice makes it entangle with private quantum key $|\phi^+\rangle_{A_i B_i}$. After the $CU_1^{dp}$ operation, $A_i$, $B_i$, and $|m_i\rangle_{dp}$ are in the state $|\Upsilon\rangle = \frac{1}{\sqrt{2}}(|0\rangle|0\rangle|0\rangle_{dp} + |1\rangle|1\rangle|1\rangle_{dp})_{A_i B_i m_i}$ or $|\Omega\rangle = \frac{1}{\sqrt{2}}(|0\rangle|0\rangle|1\rangle_{dp} - |1\rangle|1\rangle|0\rangle_{dp})_{A_i B_i m_i}$ randomly. Apparently, the traveling state $|m_i\rangle_{dp}$ in $S_M$ is always in the maximal mixture of $|0\rangle_{dp}$ and $|1\rangle_{dp}$ (i.e., $|m_i\rangle_{dp} = \frac{1}{2}|0\rangle_{dp}\,_{dp}\langle 0| + \frac{1}{2}|1\rangle_{dp}\,_{dp}\langle 1|$) during the transmission. Consequently, nothing useful will be obtained by Eve just from the ciphertext even if she intercepts it [60]. In addition, if Eve wants to decrypt out the initial state of $|m_i\rangle_{dp}$, she has to do another thing, that is, eavesdropping $S_C$ during the private quantum key sharing process. However, as analyzed above, the quantum no-cloning theorem and the security check method in the private quantum key sharing process guarantee that Eve cannot achieve her aim.

In fact, Eve cannot intercept the ciphertext without being discovered. The reason lies in two aspects [59]: one is the quantum no-cloning theorem and the other is the usage of decoy photons [81-82] in this step, which will be inevitably disturbed by Eve if she tries to intercept the ciphertext.

**Step 3:** After decrypting out the initial states of traveling states, Bob encodes his secret messages and sends the encoded traveling states to Alice. Because Eve is not aware of the initial state of $|m_i\rangle_{dp}$, she is unable to obtain $k_i$ from $U_{k_i}^{dp}|m_i\rangle_{dp}$ even if she intercepts it. In fact, due to the quantum no-cloning theorem and the usage of decoy photons [81-82] in this step, Eve cannot intercept $U_{k_i}^{dp}|m_i\rangle_{dp}$ without being detected. [59]

**Step 4:** Alice encodes her secret messages and announces the final encoded traveling state to Bob for decoding. Because there are no photons transmitted between them, Eve has no opportunity to impose an attack.

**Step 5:** For each of $|\phi^+\rangle_{AB}$, Alice and Bob perform $R(\theta)$ on their respective photon by choosing a proper angle $\theta$, respectively. At present, there are two special attack strategies toward channel-encrypting quantum cryptography protocols. The first one was suggested in Ref. [51], that is, Eve first entangles her ancilla into private quantum key $|\Phi\rangle_{AB}$ through a CNOT operation ( the traveling photon $\gamma$ is the control qubit and her ancilla is the target qubit) after intercepting $\gamma$, and then uses this entangled relation to extract something useful from key bits. The other one was suggested in Ref. [54], which always threatens QKD protocols based on reused quantum key, such as the ones of Refs. [51-52]. It has been verified in detail in Ref. [54] that, as long as $\theta \neq k\pi \pm \pi/4 \ (k=0,\pm1,\pm2,\cdots)$, $R(\theta)$ is secure and effective toward these two special attack strategies.

Note that except Step 1, the security analysis of the proposed QD protocol against Eve's active attacks is also similar to that of the one in Ref. [49], as their basic principles of dialogue process are similar.

## 4 Discussion

(1) The information-theoretical efficiency

In the Cabello's definition, the information-theoretical efficiency is $\eta = b_s / (q_t + b_t)$, [4] where $b_s$, $q_t$, and $b_t$ are the expected secret bits received, the qubits used, and the classical bits exchanged between participants, respectively. In the proposed QD protocol against collective-dephasing noise, without considering security checks, the first traveling state $|m_1\rangle_{dp}$ is used for exchanging Alice and Bob's respective one bit. In the meanwhile, three qubits are consumed for producing private quantum key $|\phi^+\rangle_{A_1B_1}$, and one classical bit is consumed for announcing the measurement outcome of $U_{j_1}^{dp}U_{k_1}^{dp}|m_1\rangle_{dp}$. Considering that private quantum key $|\phi^+\rangle_{A_1B_1}$ will always be repeatedly used in the following runs, the three qubits for producing it may be ignored. Consequently, it nearly follows that $b_s = 2$, $q_t = 2$, and $b_t = 1$. Hence, the information-theoretical efficiency of the proposed QD protocol is nearly 66.7%. The same conclusion can also be drawn under the case of collective-rotation noise.

(2) Comparisons of previous information leakage resistant QD protocols

Apparently, all the QD protocols of Refs. [43-50] only work under an ideal channel. Hence, the proposed QD protocols take advantage over them in defeating collective noise.

On the other hand, the comparisons between the proposed QD protocols and all the QD protocols in Refs. [78,80] are summarized in detail in Table 1. ①The proposed QD protocols consume two-photon states as traveling states and three-photon states for producing private quantum key. As private quantum key can always be reused, the three-photon states for producing it may be ignored. Hence, the proposed QD protocols nearly only consume two-photon states. Consequently, as for initial quantum resource, the proposed QD protocols defeat those of Ref. [78] and are nearly equivalent to those of Ref. [80]; ②as for quantum measurements, the proposed QD protocols defeat those of Ref. [78] and are equivalent to those of Ref. [80]; and ③the proposed QD protocols have the highest information-theoretical efficiency.

Table 1. Comparisons of previous information leakage resistant QD protocols against collective noise

|  | The protocols of Ref. [78] | The protocols of Ref. [80] | The proposed QD protocols |
|---|---|---|---|
| Initial quantum resource | Four-photon states | Two-photon states | Nearly two-photon states |
| Quantum measurements | Bell basis measurements | Single-photon measurements | Single-photon measurements |
| Information-theoretical efficiency | 40% | 33.3% | Nearly 66.7% |

(3) Implementation feasibility

At present, the entangled states can be produced with a practical entangled source, that is, a parametric down-conversion source with a beta barium borate crystal and a pump pulse of ultraviolet light [77]. The logical unitary operations, which do not alter the antinoise feature of logical qubits, can be accomplished with some linear optical elements such as half-wave plates and $\lambda/4$ plates. [77] The storage of photons can be achieved by optical delays in a fiber as suggested in Ref. [16]. The single-photon measurement needed can be accomplished via single-photon detector. All of these indicate that the proposed QD protocols can be implemented with current techniques in experiment.

(4) Comparisons of QKD&OTP

It is well known that the aim of QKD is to establish a sequence of random key between remote participants. If one wants to use QKD to distribute a secret, an additional OTP will be needed. Perhaps someone think that Alice and Bob may accomplish bidirectional communication over a noisy channel through QKD&OTP instead of the proposed QD protocols. In a QKD&OTP-based bidirectional communication, two participants need to do the following things subsequently [43] : first, they share a sequence of random key with the length equal to the total amount of their secrets through QKD using logical qubits as traveling states (i.e.,$\left|m_i\right\rangle_{dp}$ in the case of collective-dephasing noise and $\left|m_i\right\rangle_r$ in the case of collective-rotation noise); second, they encrypt their respective secret with random key; and finally, they publish their respective ciphertext via classical channel and decrypt out each other's secret with random key. In this case, it follows that $b_s = 2$, $q_t = 4$, and $b_t = 2$, making $\eta = 33.3\%$. It can be concluded that the proposed QD protocols have higher information-theoretical efficiency than a QKD&OTP-based bidirectional communication.

## 5 Conclusions

In all, two fault tolerant channel-encrypting QD protocols against collective-dephasing noise and collective-rotation noise are presented in this paper, respectively. The DF states, each of which is composed of two physical qubits, act as traveling states combating collective noise. The proposed QD protocols have several remarkable features:

①Two participants first share private quantum key (i.e., EPR pairs)securely over a collective-noise channel. Then, they share the initial state of each traveling two-photon logical qubit privately through encryption and decryption with private quantum key. As a result, the issue of information leakage is overcome.

②The private quantum key can be repeatedly used after rotation as long as the rotation angle is properly chosen, making the information-theoretical efficiency nearly as high as 66.7%, which is much higher than those of previous information leakage resistant QD protocols against collective noise.

③With respect to quantum measurements, only single-photon measurements rather than two-photon joint measurements are needed.

④As for communication security, an eavesdropper cannot obtain anything useful about secret messages during the dialogue process without being discovered.

⑤On the aspect of implementation feasibility, they can be implemented with current techniques in experiment.

**Acknowledgment**

Funding by the National Natural Science Foundation of China (Grant Nos. 61402407 and 11375152) is gratefully acknowledged.